# THE UNIVERSITY OF WARWICK

# Optimizing Device-to-Device Communications in Cellular Networks

**First year research report**

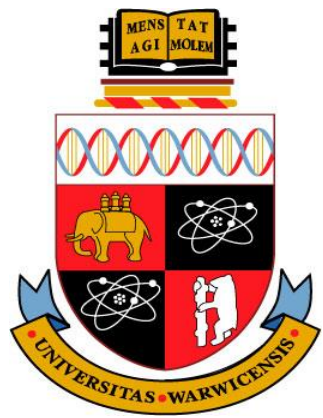

Hu YUAN

1053533

Supervisors: Dr W. GUO

Dr A. SHAH

# List of Figure

# List of table

# Table of Contents

# Abstract

In this report, we consider how to dynamically select transmission bands and multi-hop routes for device-to-device (D2D) communications in co-existence with a cellular network overlay. Firstly, we consider different wireless routing algorithms, i.e. broadcasting-routing (BR) method, and shortest-path-routing (SPR) method. The results show that depending on the co-existence cellular users' outage constraint, different routing strategies have different merits. BR is acceptable at the low D2D user density but is terrible at high density.

We also consider the channel band performance (Uplink band and Downlink band). The results show that the multi-hop D2D can achieve a low outage probability using the uplink band (approximately 5%), and D2D in the downlink band performs a little poorly (approximately 12%) outage with SPR.

# 1. Introduction

Recently, cellular networks already face the problem of capacity or resources limitations due to increasing demands of various types of communications. For solving those problems, wireless local networks (WLANs) have become exceedingly popular, as they enable access to the Internet and local services easily at a low price and high fast access to network. However, the coverage of the WLANs is limited and it operates on a license exempt bands. This article introduces a technology component for LTE Advanced: device-to-device (D2D) communication as an underlay to cellular networks.

The concept of D2D communication is an underlay to a cellular network, operating potentially on the same resources [1]. The cellular network operates in licensed bands, and it is important to guarantee that D2D transmissions will not generate harmful interference to cellular users [2]. This concept is illustrated in Figure 1, where user equipment (UE) is served by the network via the base stations, UE units could communicate directly with each other over the D2D links. The UE in D2D connections remains controlled by the base station and continue cellular operation.

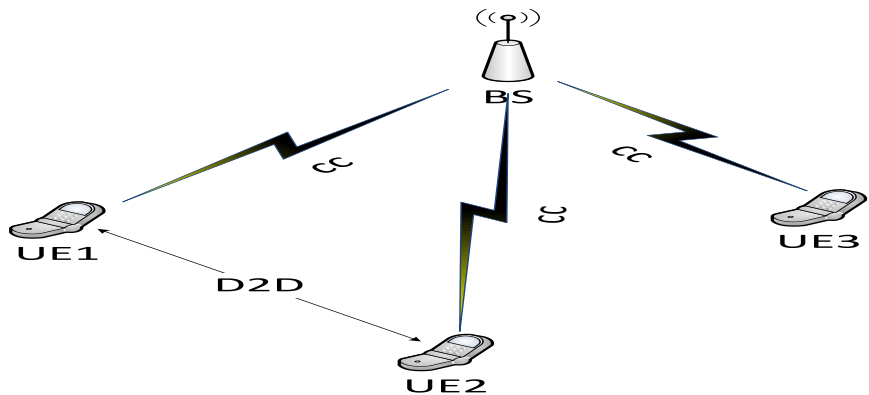

**Figure 1 Device to Device communication between UE1 and UE2 as an underlay to a cellular network. CC is a Conventional Cellular communication channel.**

Compared to other local connectivity solutions, Bluetooth or WLAN, the D2D communication supported by a cellular network offers additional compelling advantages: [2]

Firstly, the network can advertise local services available within the current cell. Thus for automated service discovery, the devices do not have to constantly scan for available WLAN AP or Bluetooth devices.

Secondly, the cellular network can be established without manual pairing of devices.

Thirdly, the D2D communication under control of the Base Stations (BSs) for better interference management and mitigation

Several wireless standards have addressed the need for D2D operation in the same band as the BSs, such as Hiperlan 2 [8], TETRA [9], and WLAN. In all these standards D2D communication is assumed to occur on separate resources.

## 2. Literature review

D2D communications could bring great potential benefits to system capacity and spectral efficiency [10]; however it might cause severe interference to the cellular UEs. Thus, an efficient interference management mechanism and routing selection should be developed to guarantee a target level of performance for cellular UEs.

Interference aware wireless routing selection has been studied recently in [14] for a wireless multi-hop network. It has been shown that when interference is taken into account, the relay selection choice and performance is significantly different to that of noise-limited networks [15] [16]. There have been a limited number of studies on the effect of mutual interference between cellular networks and D2D networks in co-existence

In [11], a D2D communications is assisted multi-input multi-output (MIMO) downlink broadcast channel in the cellular networks, which could offload the overhead of signalling and retransmission at the BSs. At the same time, the UEs with limited decoding capabilities could be enhanced by user cooperation within D2D group. However, the interference to the Conventional Cellular (CC) users is significant when the D2D are communicating with MIMO. Additionally, the MIMO system for the D2D device is complicated that cannot be shared with CC users.

An interference constrained relay selection of D2D communication is issued in [12]. In this, the transmission rate from D2D UE is analysed for constrained the interference. It deduces a relay selection rule based on the interference constraining, but which does not take into account the complex geometry of a multi-cell cellular network.

In [13], the D2D is set as pair to maximize the number of permitted D2D communication pairs in a system meanwhile avoiding the strong interference from D2D communication to the cellular communication. However, in an actual environment the number of the D2D communication pair is dynamically and random.

In [17], the authors consider how cooperative partner selection can be optimised in a random user positioning environment. Deterministic expressions are derived for the optimal number of cooperative partners as a function of the channel strengths.

# 3. System Setup

## 3.1. System Topology

The system is an OFDMA based downlink multiple-access network such as 4G LTE. It consists of 19 static macro base-stations (BSs) and UEs [18].

This report considers 3 different transmission modes in co-existence:

1) Conventional Cellular (CC): the source UE transmits data to the serving Base station (BS) using the uplink band and the destination UE receives data from the same or different serving BS in the downlink band.
2) D2D Communications in Uplink Band (D2D-UL): the source UE transmits data to the relaying and the destination UEs using the uplink band and the interference at each UE is from other UEs, shown as Figure 2 (a).
3) D2D Communications in Downlink Band (D2D-DL): the source UE transmits data to the relaying and the destination UEs using the downlink band and the interference at each UE is from BSs (this may or may not include the parent BS), shown as Figure 2 (b).

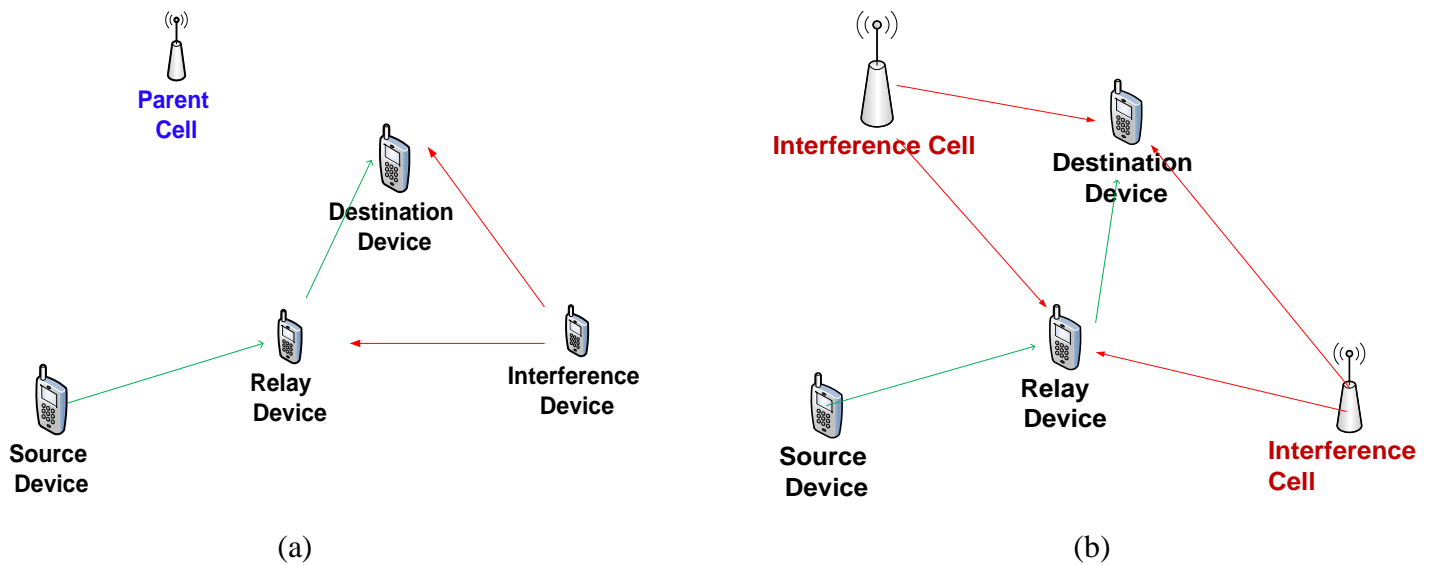

(a)     (b)

**Figure 2 D2D communications using cellular uplink band (D2D-UL) (a), D2D communications using cellular downlink band (D2D-DL) (b). The red lines indicate the interference and green lines indicate the D2D communication.**

This report assumes D2D communication links occur in the same frequency band as CC transmissions, and there is mutual interference within a BS coverage region. It also assumes that the traffic model for both CC and D2D is full buffer, meaning that there is always data transmitted on all frequency bands. The D2D transmissions employ a multi-hop Decode-and-Forward (DF) mechanism with no cooperative combining at the receiver.

## 3.2. System Metrics

The instantaneous signal-to-noise-interference ratio (SINR) of a communication link from *n* to *m* is defined as:

$$\gamma_{n,m} = \frac{H_{n,m} P_{n,m} K r_{n.m}^{-\alpha}}{\sigma^2 + \sum_{\substack{i \in \Phi \\ i \neq n}} H_{i,m} P_{i,m} K r_{i.m}^{-\alpha}} \quad (1)$$

Wher $\sigma^2$ is the AWGN power, *H* is the fading gain, *P* is the transmit power, *K* is the frequency dependent path loss, and *r* is the distance. Define $\beta$ as the parameter of the exponential distribution $G = HP_{n,m}K$ There is a set of $\Phi$ interferers, and it can be assumed that for an interference-limited network, the AWGN power is negligible compared to the sum of interference powers..

For the down link channel, the interference signal is from other co-channel base stations and for uplink the interference is from the other co-channel UEs in other cells. Figure 3 shows the CDF of SINR in an example cell.

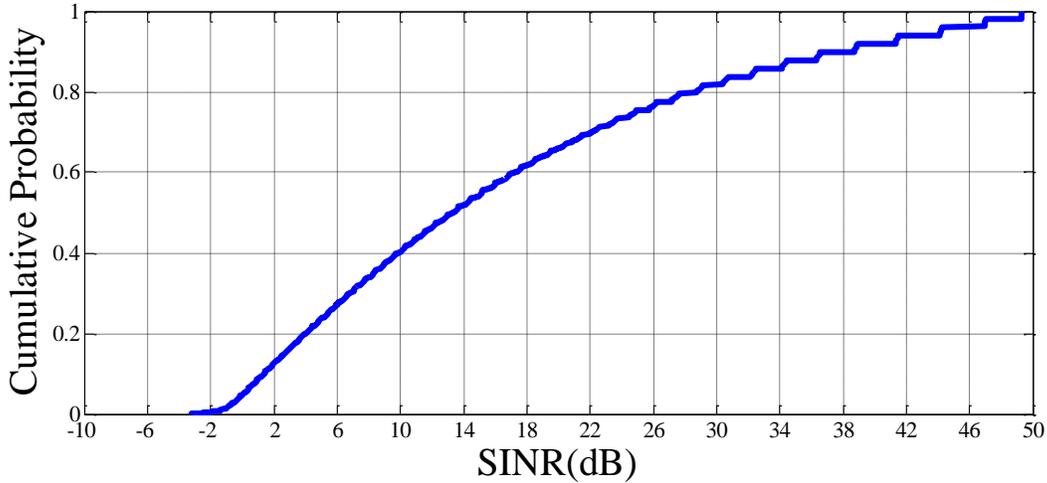

**Figure 3 shows the CDF of SINR in the central cell of Macro BSs.**

The single link outage probability is defined as the probability that the average SINR ($\bar{\gamma}$) is lower than a threshold, where for 4G LTE, the typical value of is -6 dB, below this value is not possible [3].

The average of multiple links' outage probability is defined as $P_{X-Y,out}^{Z}$, where *X* refers to the communication links (D2D or CC), *Y* refers to the scheme used and *Z* indicates the transmission bands. The following sub-scripts are used to denote transceiver nodes:

- Macro BSs are denoted by sub-script $n \in N$.

- Mobile UEs are denoted by sub-script $m \in M$.
- Interference nodes are denoted by sub-script $i \in \Phi$

This report now considers in detail the outage probabilities for CC and D2D communications.

**Table 1. LTE Systerm Parameters**

| Parameter | Symbol | Value |
|---|---|---|
| Bandwidth | | 20 MHz |
| Frequency | | 2.1 GHz |
| Data Connectivity Threshold | $\zeta$ | -6 dB |
| CC UEs per BS | | 1 on each band |
| CC UE density per BS | $\Lambda_{CC}$ | 1.27 per km$^2$ |
| D2D UEs per BS | | 10–140 (1 on each band) |
| D2D sender node density | $\Lambda_{DD}$ | 1.27 per km$^2$ |
| Macro-BS coverage radius | $R_{BS}$ | 500 m |
| Macro-BS density | $\Lambda_{BS}$ | 1.27 per km$^2$ |
| Propagation model | | [11] |
| AWGN power | $\sigma^2$ | $6 \times 10^{-17}$ W |
| BS-UE antenna height diff. | | 35 m |
| BS antenna | | Omni-directional |
| BS transmit power | $P_{BS}$ | 40 W |
| D2D transmit power | $P_{D2D}$ | 0.1 W |
| CC UE transmit power | $P_{CC}$ | 0.1 W |
| Traffic model | | Full Buffer |
| No. Macro-BS | | 19 (wrap around) |
| Antenna Scheme | | SISO |
| Multipath fading | | Rayleigh |
| Shadow fading variance | | 6 dB |

## 4. Channel Selection for D2D Communication

### 4.1. Conventional Cellular (CC) Communication

The conventional cellar communication involves UEs communicating through serving BSs. A system is shown in Figure 4, where all nodes (BSs and UEs) are distributed randomly and uniformly. Assuming that the closest BS to a UE is the serving BS, the PDF of the distance r between UE to BS can be derived from a 2-D Poisson process [4] [5] [6]:

$$f_R(r) = 2\Lambda\pi r \exp(-\Lambda\pi r^2) \qquad (2)$$

where $\Lambda$ is the transmit node density.

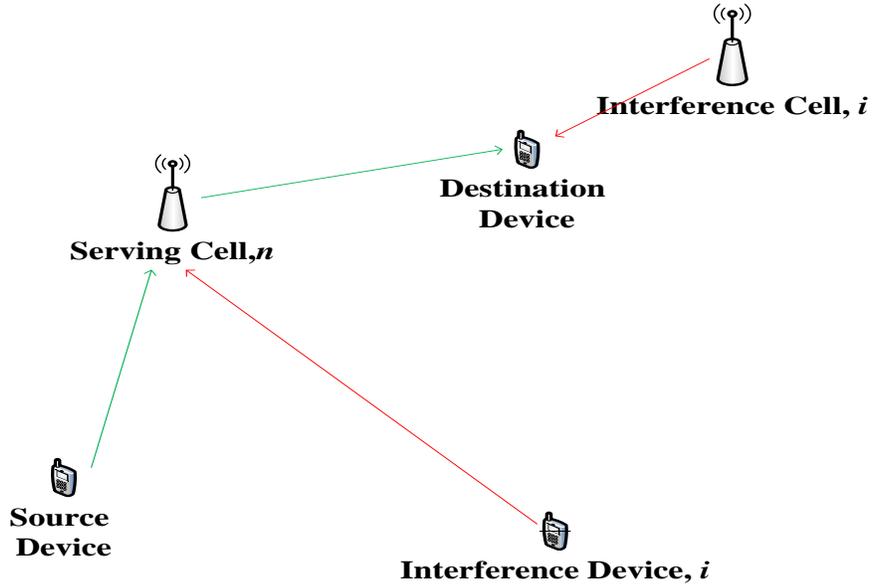

**Figure 4 Conventional cellar communications.**

The analysis considers 2 arbitrary UEs, which have a distance of $r_{m,n}$ and $r_{n,m'}$ to their respective serving BS. For CC communications, the outage probability of UE *m* communicating to UE *m'* is given by the difference between one and the product of the uplink and downlink coverage probabilities:

$$P_{CC,out}(m,m') = 1 - \mathrm{P}(\bar{\gamma}_{m,n} > \zeta)\mathrm{P}(\bar{\gamma}_{n,m'} > \zeta). \tag{3}$$

We consider a network with $\Lambda_{BS}$ deployed co-frequency BSs, where each BS has a certain number of UEs. For downlink transmission, the interference arrives from adjacent BSs. For uplink transmission, the interference arrives from other UEs in adjacent BSs, and there is no interference from uplink UEs within the BS.

Stochastic geometry used to derive analytical expressions for the outage probability in presence of Rayleigh fading and employs numerical simulation to verify the results. The probability of successful transmission in the uplink and is: [16]

$$\mathrm{P}(\bar{\gamma}_{m,n} > \zeta) = \exp\left[-\Lambda_{BS} \pi r_{m,n}^2 \mathrm{A}(\zeta,\alpha)\right] \tag{4}$$

The probability of successful transmission in the downlink is: [16]

$$\mathrm{P}(\bar{\gamma}_{n,m'} > \zeta) = \exp\left[-\Lambda_{BS} \pi r_{n,m'}^2 \mathrm{A}(\zeta,\alpha)\right] \tag{5}$$

Where the A( ) function is given by:

$$A(\zeta,\alpha) = \int_{\zeta^{-2/\alpha}}^{+\infty} \frac{\zeta^{2/\alpha}}{1+u^{\alpha/2}} du,$$

$$= \sqrt{\zeta} \arctan(\sqrt{\zeta}) \text{ for } \alpha = 4. \tag{6}$$

Therefor the outage probability of CC communication is:

$$P_{CC,out}(m,m') = 1 - \exp\left[-\Lambda_{BS}\pi(r_{m,n}^2 + r_{n,m'}^2)A(\zeta,4)\right] \tag{7}$$

This is the network outage probability. The result for the theoretical expression and simulation is shown in Figure 5.

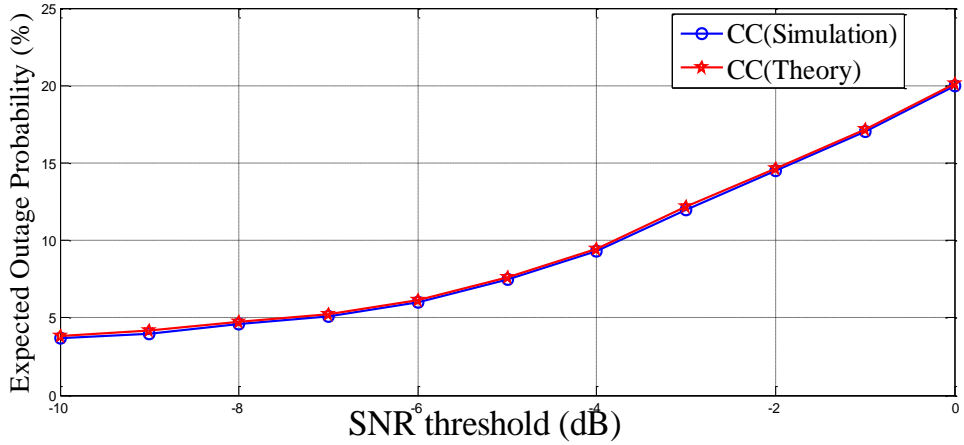

**Figure 5 CC communications network outage probability for a multi-BS network simulation and theoretical results.**

### 4.2. Downlink Channel (DL) Communication

D2D Communications in Downlink Band (D2D-DL): the source UE transmits data to the relaying and the destination UEs using the downlink band and the interference at each UE is from BSs.

The outage probability is:

$$E_R\left(P_{D2D,out}^{DL}\right) = 1 - \frac{\alpha \sin(2\pi/\alpha)}{2\pi[1+A(\zeta)]} \frac{\zeta^{-2/\alpha}}{1+\Lambda_{BS}/\Lambda_{DD}} \tag{8}$$

Where $\alpha$ path loss parameter, $\zeta$ is Data Connectivity Threshold, $\Lambda_{BS}$ $\Lambda_{DD}$ are the density of base station and D2D sender node density respectively. The result is shown at Figure 6.

### 4.3. Uplink Channel (UL) Communication

D2D Communications in Uplink Band (D2D-UL): the source UE transmits data to the relaying and the destination UEs using the uplink band and the interference at each UE is from other UEs.

The outage probability is:

$$E_R\left(P_{D2D,out}^{DL}\right) = 1 - \frac{\alpha \sin(2\pi/\alpha)}{2\pi[1+A(\zeta)]} \frac{\zeta^{-2/\alpha}}{1+(\Lambda_{BS}/\Lambda_{DD})(P_{BS}/P_{D2D})^{2/\alpha}} \quad (9)$$

Where $\alpha$ path loss parameter, $\zeta$ is Data Connectivity Threshold, $\Lambda_{BS}$ $\Lambda_{DD}$ are the density of base station and D2D sender node density respectively. P is the transmitter power. The result is shown at Figure 6.

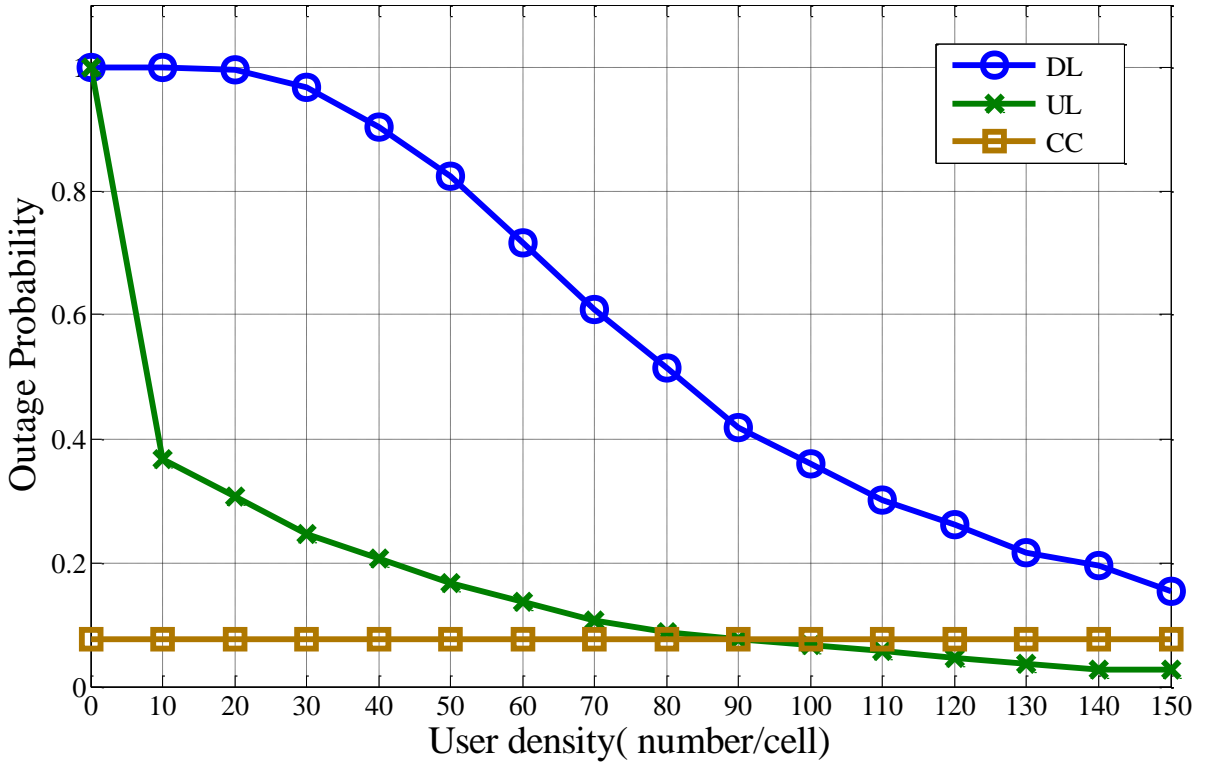

**Figure 6 The simulated outage of probability of D2D communications with different channels at different density.**

## 5. Route Selection for D2D Communication

### 5.1. Broadcast Routing (BR)

#### 5.1.1. Algorithm of BR

In BR, the transmitter knows the identify information of receiver though the BSs. the step-by-step D2D algorithm needed to achieve shortest path routing from UE *m* to UE *m'*:

### 5.1.2. The outage of BR

For D2D communications, this report considers additional UEs that cannot be scheduled radio resources to transmit their data. Therefore, UEs with delay-tolerant data has been allocated a D2D channel to communicate with each other. The caveat is that the communication occurs in the same frequency band as the parent BS and there is now interference from CC onto the co-existing D2D links and vice versa. In any transmission band, the outage probability for DF relaying is given as a function of the product of the success probability for each link:

$$P_{D2D,out} = 1 - \prod_{j=1}^{J}\left(1 - E_R\left(P_{D2D\text{-}BR,out}\right)\right) \qquad (10)$$

where the total number of hops *J* is determined by the density of UEs in the network, the distance between the source and destination UEs, and the route selected. The Figure 9 shows the simulation result of the outage probability of the BR D2D communication. The performance is well when the user density is low but it cause significant interference.

## 5.2. Shortest Path Routing (SPR)

### 5.2.1. Algorithm of SPR

In SPR, each D2D UE node knows its location through GPS and other wireless transmission means. This report now outlines the step-by-step D2D algorithm needed to achieve shortest path routing from UE *m* to UE *m'*:

Figure 7 illustrates this process using downlink transmission bands.

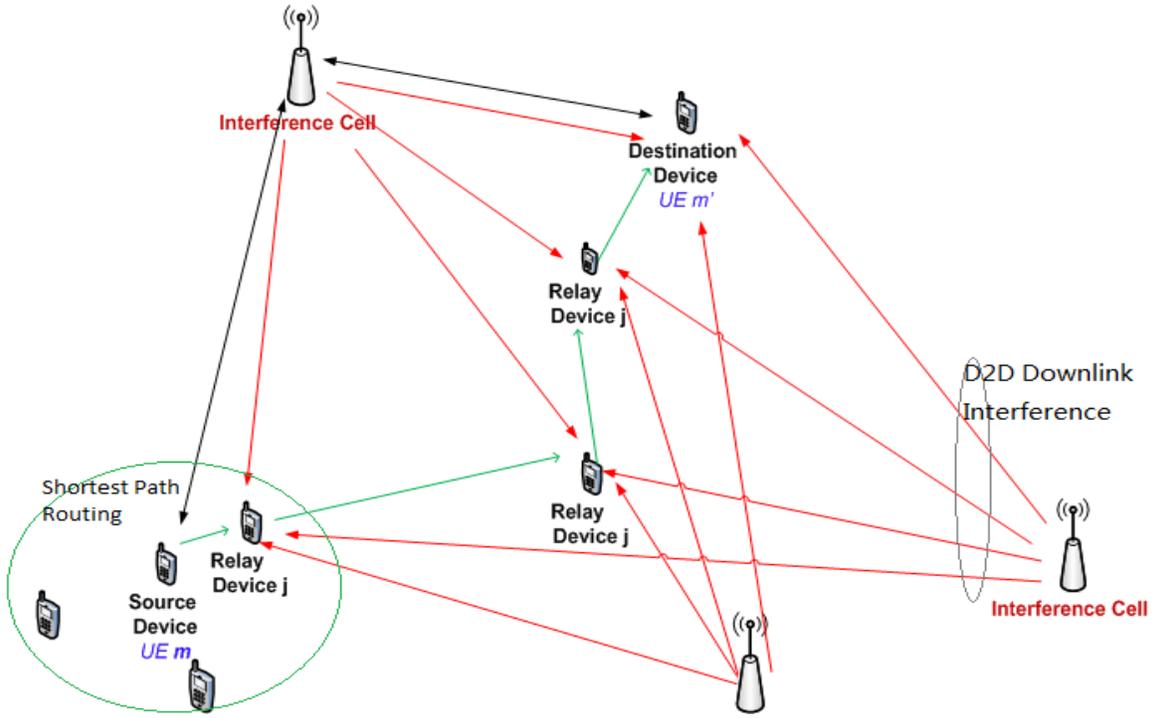

**Figure 7 Illustration of shortest path routing (SPR) between UE *m* and *m'*.**

### 5.2.2. Outage Probability

For the UL transmission band, the D2D and CC transmissions can be thought of as a 2-tier heterogeneous network, where all interfering transmissions are transmitted with the same uplink UE power. The expected D2D outage probability for a single link, averaged across all distances is:

$$E_R\left(P^{UL}_{D2D-SPR,out}\right) = 1 - \frac{N_{D2D} - 1}{N_{D2D} - 1 + 2B(\zeta, \alpha)} \quad (11)$$

Where B ( ) function is given by:

$$B(\zeta, \alpha) = \int_0^{+\infty} \frac{\zeta^{2/\alpha}}{1 + u^{\alpha/2}} du$$

$$= \frac{\pi\sqrt{\zeta}}{2} \quad \text{for } \alpha = 4. \quad (12)$$

The Figure 9 is shown the outage probability of the D2D communication with SPR.

## 6. Conclusion

In this report, we consider how to dynamically select transmission strategies for D2D communications in co-existence with a conventional cellular (CC) overlay. The Monte-Carlo simulations and stochastic geometry are used to devise an algorithm for selecting the transmission band and D2D routing path method. The benefit of this algorithm is that a serving base station is able to make decisions that balance the performance of both regular cellular users and D2D users.

## Reference

[1] Klaus Doppler, Mika Rinne, Carl Wijting, Cássio B. Ribeiro, and Klaus Hugl 'Device-to-Device Communication as an Underlay to LTE-Advanced Networks' *IEEE Communications Magazine*, Volume:47 , Issue: 12, Page(s):42 – 49, Dec. 2009.

[2] P. J¨anis, C.-H. Yu, K.Dopper, C. Ribeiro, C. Wijting, K. Hugl, O. Tirkkonen, and V. Koivunen, 'Device-to-Device Communication Underlaying Cellular Communications Systems '*Int. J. Communications, Network and System Sciences*, Pages169-247, March 2009.

[3] 3GPP, "TR36.814 V9.0.0: Further Advancements for E-UTRA Physical Layer Aspects (Release 9)," *3GPP, Technical Report*, Mar. 2010.

[4] M. Haenggi, J. G. Andrews, F. Baccelli, O. Dousse, and M. Franceschetti, "Stochastic geometry and random graphs for the analysis and design of wireless networks," in *IEEE Journal on Selected Areas in Communications* (JSAC), vol. 28, pp. 1029–1046 , Sep. 2009.

[5] M. Haenggi, Stochastic Geometry for Wireless Networks. England, UK: Cambridge University Press, 2012.

[6] Z. Gong and M. Haenggi, "Interference and Outage in Mobile Random Networks: Expectation, Distribution, and Correlation," in *Transactions on Mobile Computing, IEEE*, vol. 11, Dec. 2012.

[7] S. Wang, W. Guo, and T. O'Farrell, "Two-Tier Cellular Networks with Frequency Selective Surface," in *IEEE International Conference on High Performance Computing and Communications* (HPCC), Jun. 2012.

[8] ETSI, "BRAN; HIPERLAN2 type 2; data link control (DLC) layer; part 4: Extension for home environment," TS 101 761-4, v1.3.2, 2002.

[9] ETSI, "Terrestrial trunked radio (TETRA); voice plus data (V+D) designers' guide; part 3: Direct mode operation (DMO)," TR 102 300-3 v1.2.1, 2002.

[10] Fodor, G. Dahlman, E. ; Mildh, G. ; Parkvall, S. ; Reider, N. ; Miklós, G. ; Turányi, Z. 'Design aspects of network assisted device-to-device communications' *Communications Magazine, IEEE* ,Volume:50 , Issue: 3 , Page(s):170 – 177, March 2012.

[11] Chun-Che Chien, Hsuan-Jung Su ; Hsueh-Jyh Li, 'Device-to-Device assisted downlink broadcast channel in cellular networks' *Wireless Personal Multimedia Communications (WPMC), 2012 15th International Symposium on*, Page(s):85 – 89, 24-27 Sept. 2012.

[12] L. Wang, T. Peng, Y. Yang, and W. Wang, "Interference constrained relay selection of D2D communication for relay purpose underlaying cellular networks," in *IEEE Wireless Communications, Networking and Mobile Computing Conference*, pp. 1–5, Sep. 2012.

[13] Yanfang Xu ; Rui Yin ; Tao Han ; Guanding Yu, 'Interference-aware channel allocation for Device-to-Device communication underlaying cellular networks', *IEEE International Conference on Communications in China: Wireless Communication Systems (WCS),* Page(s): 422 – 427, 15-17 Aug. 2012.

[14] G. Parissidis, M. Karaliopoulos, T. Spyropoulos, and B. Plattner, "Interference-aware routing in wirless multihop networks," *in IEEE Transactions on Mobile Computing*, vol. 10, pp. 716–734, May 2011.

[15] W. Guo and T. O'Farrell, "Relay deployment in cellular networks: Planning and optimization," in *IEEE Journal on Selected Areas in Communications (JSAC)*, vol. 30, Nov. 2012.

[16] G. Parissidis, M. Karaliopoulos, T. Spyropoulos, and B. Plattner, "Maxmin relay selection for legacy amplify-and-forward systems with interference," in *IEEE Transactions on Wireless Communications*, vol. 8, pp. 3016–3027, Jun. 2009.

[17] W. Guo, I. Chatzigeorgiou, I. J. Wassell, R. Carrasco "Partner selection and power control for asymmetrical collaborative networks", in *IEEE Vehicular Technology Conference (VTC 2010-Spring),* pp. 1-5, May 2010.

[18] W. Guo, S. Wang, X. Chu, J. Zhang, J. Chen, H. Song "Automated small-cell deployment for heterogeneous cellular networks", in *IEEE Communications Magazine,* vol. 51, pp. 46-53, May 2013.